\documentclass[prd,twocolumn,showpacs,letterpaper,nofootinbib]{revtex4}
\usepackage{graphics}
\usepackage{epsfig} 
\usepackage{amsmath}

\input epsf

\newcommand{\be}{\begin{equation}}
\newcommand{\ee}{\end{equation}}
\newcommand{\bea}{\begin{eqnarray}}
\newcommand{\eea}{\end{eqnarray}}
\def\pslash{{\cal P}{\hbox{\kern-6pt $\slash$}}}

\long\def\comment#1{}

\input{psfig.tex}
\begin{document}
\title{Expansion, Geometry, and Gravity}
\author{Robert R. Caldwell}
\affiliation{Department of Physics \& Astronomy, Dartmouth College,
6127 Wilder Laboratory,
Hanover, NH 03755}
\author{Marc Kamionkowski}
\affiliation{Mail Code 130-33, California Institute of Technology,
Pasadena, CA 91125}

\date{\today}

\begin{abstract}
In general-relativistic cosmological models, the expansion history, matter
content, and geometry are closely intertwined.  In this brief paper, we clarify
the distinction between the effects of geometry and expansion history on the
luminosity distance. We show that the cubic correction to the Hubble law,
measured recently with high-redshift supernovae, is the first cosmological
measurement, apart from the cosmic microwave background, that probes directly
the effects of spatial curvature. We illustrate the distinction between
geometry and expansion with a toy model for which the supernova results already
indicate a curvature radius larger than the Hubble distance.
\end{abstract}

\pacs{}
\maketitle
\vskip 1cm

\smallskip

Perhaps the most surprising implication of the general-relativistic
cosmological models which first emerged roughly eighty years ago
\cite{moderncosmology} was the possibility that the spatial geometry of an
isotropic homogeneous Universe could be nontrivial.  Alternatives to the
intuitively obvious Euclidean geometry (a ``flat'' Universe) include a
three-sphere (the ``closed'' Universe) and a three-hyperboloid (the ``open''
Universe).

Over the past five years, a suite of cosmic microwave background (CMB)
experiments have provided unique tests of the spatial geometry and provided
very compelling  evidence that the Universe is flat \cite{kss,cmbexperiments}.
More precisely, the evidence is that the Universe is {\it consistent} with
flat, by which we mean that if the Universe is open or closed, the radius of
curvature is larger than roughly three times the Hubble distance. 

As robust as the CMB results may be, the cosmological geometry is of such
fundamental significance that it is well worth exploring alternative avenues.
Inspired by the recent announcement of the measurement with high-redshift
supernovae \cite{newsupernova} indicating a transition from cosmic deceleration
to acceleration, we here point out that the  cubic correction to Hubble's
linear relation between the luminosity distance and the cosmological redshift
depends on the spatial geometry \cite{chiba,visser}.  Although well appreciated
by experts, the distinction between the effects of expansion and geometry is
often imprecise in the literature. This probably stems from the fact that the
matter content, expansion, and geometry are linked in general relativity.  In
particular,  previous measurements that found an accelerating expansion
\cite{sne} have not yet directly probed the spatial curvature. We argue here
that {\it the measurement of the cubic correction to the
luminosity-distance--redshift relation is the very first non-CMB cosmological
test that probes the spatial geometry}.   

Below we first clarify the dependence of the luminosity distance on the
geometry and on the expansion history. We then discuss a straw-man model (that
respects general relativity but postulates an exotic form of matter) that
illustrates the effects of geometry and expansion history on the cubic term.
Next we show how the the quartic correction to the Hubble law might then
distinguish between geometry and expansion history in this toy model. We then
argue that no other classical (non-CMB) cosmological tests have yet probed the
geometry.

The luminosity distance of an object at redshift $z$ is $d_L = (L/4\pi
F)^{1/2}$, where $L$ is the luminosity for a given object (presumed known for
the high-redshift supernovae being detected), and $F$ is the energy flux
received from that object. The expression for the luminosity distance in a
dynamic, homogeneous, isotropic spacetime is
\begin{equation}
  {d_L(z)\over (1+z)} = \left\{ \begin{matrix}
         R \sinh \left[ {c\over H_0 R}
     \int_0^z {dz' \over E(z')}\right], & {\rm open},\\
     c H_0^{-1} \int_0^z {dz' \over E(z')}, & \qquad {\rm flat},\\
     R \sin \left[ {c\over H_0 R}
     \int_0^z {dz' \over E(z')}\right], & {\rm closed}.\\
\end{matrix} \right.
\label{eqn:yequation}
\end{equation}
Here, $H_0$ is the Hubble constant, $c H_0^{-1}\simeq 4300$ Mpc is the Hubble
distance, and $R$ is the (comoving) radius of curvature of the open or closed
Universe; for the closed Universe, it is precisely the radius of the
three-sphere, and the argument of the sine is the angle around the three-sphere
subtended by an object at redshift $z$.  Note that as $R\rightarrow\infty$, the
open/closed expressions reduce to that for the flat Universe, as they should. 
The explicit appearance of $R$ in these equations, as well as the sin and the
sinh in the closed- and open-Universe expressions, are a direct consequence of
the non-Euclidean spatial geometry.

The function $E(z)$ quantifies the expansion rate as a function of redshift; it
is defined from
\begin{equation}
     [H(z)]^2 \equiv \left( {\dot a \over a}\right)^2 = H_0^2
     [E(z)]^2,
\end{equation} 
where $H(z)$ is the expansion rate at redshift $z$, $a(t)=(1+z)^{-1}$ is the
scale factor at time $t$, and the dot denotes derivative with respect to time.
Eqs. (\ref{eqn:yequation}) follow from the properties of the spacetime,
independent of the laws of gravitation.

According to general relativity, the expansion rate is determined by the matter
content and the curvature.  If the Universe consists of nonrelativistic matter
(whose energy density scales as $a^{-3}$) plus a cosmological constant (whose
energy density remains constant), then
\begin{equation}
     E(z) = \left[ \Omega_m (1+z)^3 + \Omega_\Lambda +
     \Omega_K(1+z)^2\right]^{1/2},
\label{eqn:Eequation}
\end{equation}
where
\begin{equation}
     \Omega_m= {8 \pi G \rho_m \over 3 H_0^2}, \quad \Omega_K =
     {c^2 \over (a_0 H_0 R)^2}, \quad \Omega_\Lambda = {\Lambda c^2 \over 3 H_0^2},
\end{equation}
and $\rho_m$ is the nonrelativistic-matter density today, $G$ is the
gravitational constant, and $\Lambda$ is the cosmological constant.   For
reference, current best-fit values from a variety of measurements are
$\Omega_m\simeq0.3$ and $\Omega_\Lambda\simeq0.7$.

Eq. (\ref{eqn:Eequation}) implies $\Omega_m+\Omega_\Lambda+\Omega_K=1$. Thus,
measurement of $\Omega_m$ and $\Omega_\Lambda$ (i.e., determining the matter
content of the Universe) fixes $\Omega_K$, even without observing any
geometrical effects.  This is the explicit statement that in general relativity
the matter content of the Universe determines the geometry.  

Let us now return to Eq. (\ref{eqn:yequation}).  This can be Taylor expanded
(see, e.g., Ref. \cite{visser}),
\begin{widetext}
\begin{equation}
     d_L(z) = {cz \over H_0} \left\{ 1+{1\over 2} (1-q_0)z -
     {1\over 6}\left[1-q_0-3q_0^2+j_0\pm{c^2\over H_0^2 R^2}
     \right] z^2+{\cal O}(z^3) \right\},
\label{eqn:central}
\end{equation}
\end{widetext}
where $\pm$ is $+$ for a closed Universe and $-$ for an open Universe.  In this
expression, $q_0\equiv-(\ddot a/a)H_0^{-2}$ (evaluated today) is the well-known
deceleration parameter, and $j_0 \equiv (\dddot a /a)H_0^{-3}$ (also evaluated
today) is the ``jerk'', a scaled third time derivative of the scale factor
$a(t)$ of the Universe.  The linear term ($H_0$) in this relation is the Hubble
law, and the quadratic correction ($q_0$) is the deceleration.  These do {\it
not} depend on $R$; they thus both depend only on the expansion history, but
not on the spatial curvature.

\begin{figure*}[t] 
\includegraphics[scale=0.7]{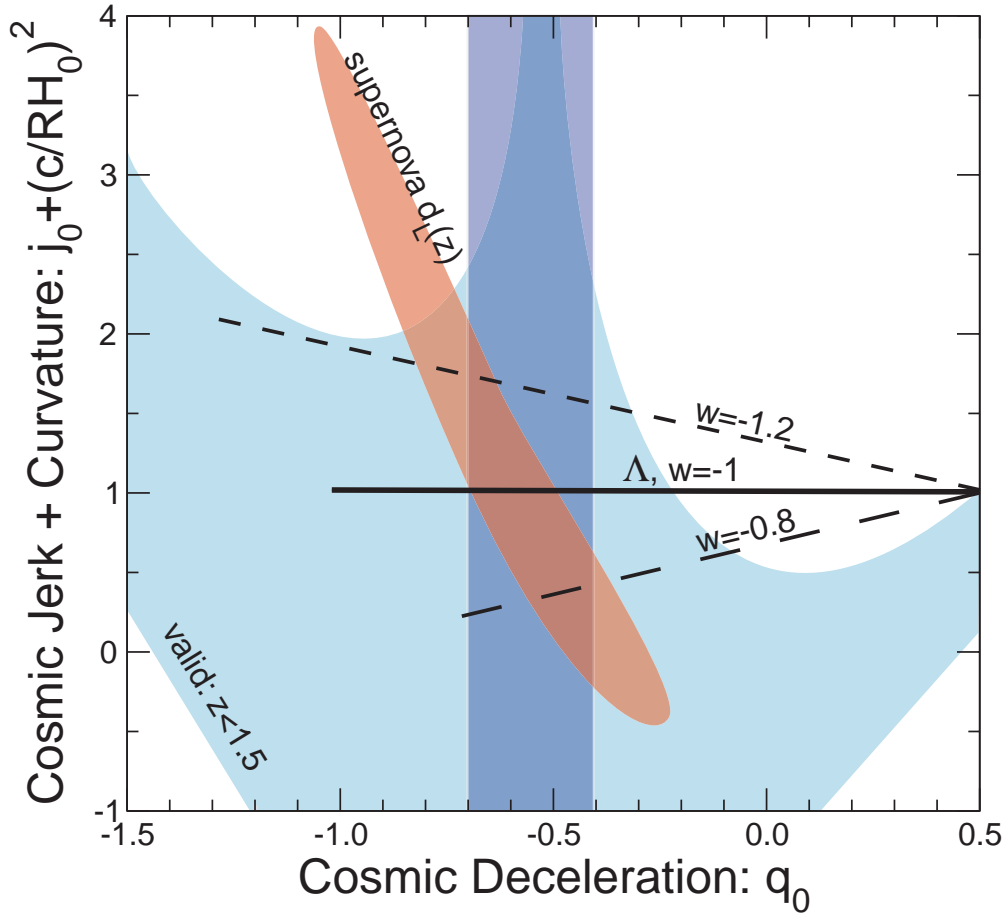}
\caption{Current constraints to the $[q_0,\, j_0+(H_0 R)^{-2}]$ plane. The dark
     shaded region is the $95\%$ confidence-level constraint from recent
     high-redshift supernova measurements \protect\cite{newsupernova}.  The
     light-shaded region shows the domain of validity of the cubic redshift
     expansion; more precisely, outside these regions, there would be a unit
     magnitude error at $z=1.5$ introduced by the quartic term.    The solid
     curve indicates a family of flat cosmological-constant models with
     decreasing matter density from right to left, terminating at $q_0=-1$ when
     $\Omega_m=0$.  The short-dash curve shows the same for flat models with
     quintessence with $w=-1.2$ (i.e., a phantom-energy model
     \protect\cite{rrc,Caldwell:2003vq}), and the long-dash curve shows the
     same for $w=-0.8$.  The vertical band shows the range of values for the
     spatially-curved model with $\Omega_m+\Omega_\Lambda=1$ discussed in the
     text and matter density spanning the range $0.2 < \Omega_m < 0.4$.}
\label{figure}
\end{figure*}

The cubic term in Eq. (\ref{eqn:central}) is the first term in the Taylor
expansion that depends explicitly on the curvature $R$.  In other words, it is
the first term in which the dependence of the brightness of an object depends
on the convergence or divergence of nearly parallel light rays that occurs in a
non-Euclidean spatial geometry. This is the first key point of our paper.

In a Universe that consists of nonrelativistic matter and a cosmological
constant, the densities $\Omega_m$ and $\Omega_\Lambda$ determine the
deceleration parameter and they determine the geometry; it has thus become
quite common for cosmologists to associate the cosmological expansion with the
cosmological geometry.  The second key point of our paper is to remark that
there is indeed a distinction. For example, if the expansion dynamics were
determined by some theory other than general relativity---as some have begun to
speculate in view of the apparent cosmic acceleration---then the geometry would
not necessarily be fixed by the expansion history.  Even within the context of
general relativity, they need not be linked, as the following toy model
illustrates.

Consider a family of models that contain, in addition to nonrelativistic matter
and a cosmological constant, some form of exotic matter with an energy density
that scales as $a^{-2}$ with the scale factor $a(t)$; this could be
accomplished, for example in the single-texture model \cite{texture} or with a
network of non-intersecting strings \cite{strings} although the specific
mechanism is irrelevant here.  Suppose also that the Universe is closed. Then
the Friedmann equation becomes
\begin{equation}
   \left[ E(z)\right]^2 = \Omega_m (1+z)^3 + \Omega_\Lambda -
   \Omega_K (1+z)^2 + \Omega_t(1+z)^2,
\end{equation}
where $\Omega_t = 8\pi G \rho_t/3 H_0^2$ is the texture energy density today. 
Suppose further that the texture density is chosen so that $\Omega_t=\Omega_K$,
or equivalently, $\rho_t = 3 c^2/8\pi G (a_0 R)^2$. In this case, the texture
and curvature terms in the Friedmann equation cancel out. We thus arrive at a
family of models that have the same $\Omega_m$ and $\Omega_\Lambda$, with
$\Omega_m+\Omega_\Lambda=1$, and  identical expansion histories, but are
parameterized by a non-zero curvature $R$ that we can freely dial without
affecting $E(z)$.

If the Universe consists of nonrelativistic matter and a cosmological constant,
then general relativity gives $q_0=\frac{1}{2}(\Omega_m - 2 \Omega_\Lambda)$
and $j_0=\Omega_m  + \Omega_\Lambda$. Suppose then (consider hypothetically a
CMB-data--free world), that the matter density is determined by dynamical
measurements.  Then $q_0$ would give us $\Omega_\Lambda$, from which we could
predict $j_0$. Since the cubic term depends on the combination $j_0+c^2/(H_0
R_0)^{2}$ [cf. Eq. (\ref{eqn:central})], measurement of the cubic correction to
the Hubble law then provides a measurement of the geometry.  This is
illustrated in Fig. \ref{figure}, where current supernova constraints to the
$[q_0,\, j_0+c^2/(H_0 R_0)^{2}]$ parameter space are shown. Within the class of
our  toy models with $0.2<\Omega_m<0.4$ and $\Omega_m+\Omega_\Lambda=1$ and $R$
variable, it is clear that $R$ is restricted to be bigger than roughly the
Hubble radius.

Suppose, however, that rather than a cosmological constant, the dark energy has
some time variation \cite{quintessence} with an energy-density decay
parameterized by an equation-of-state parameter $w$.  If so, then the jerk
$j_0$ depends on $w$ as well as $\Omega_m$ and $\Omega_\Lambda$
\cite{chiba,visser}, and it thus cannot be predicted from measurements of the
deceleration parameter; i.e., the jerk and geometry cannot be distinguished by
the cubic correction to the luminosity distance in this broader class of
models. This is illustrated by the solid, long-dashed, and short-dashed lines
in Fig. \ref{figure} that indicate families of flat models with $w=-1$
(cosmological constant), $w=-0.8$ (quintessence), and $w=-1.2$ (phantom energy
\cite{rrc,Caldwell:2003vq}). In this case, however, it may be possible to go
the quartic correction to the luminosity distance \cite{visser},
\begin{widetext}
\begin{equation}
     {c z^4 \over 24\,H_0} \left[ 2-2 q_0 -15 q_0^2 -15 q_0^3 +
     5 j_0 +10 q_0 j_0+s_0 + {2c^2(1+3q_0) \over H_0^2 R^2}
     \right],
\end{equation}
\end{widetext}
where $s_0=(\ddddot a/a)H_0^{-4}$ is the ``snap'' to isolate the curvature. 
More generally, as the parameter space of cosmological models is expanded, more
terms in the expansion must be determined to isolate the effects of the
geometry. High redshift is not necessarily required, if precision measurements
at lower redshifts can be made. 

Before closing, we note that the recent supernova \cite{newsupernova}
measurement of the cubic correction to the luminosity distance is the first
non-CMB empirical probe of the spatial geometry of the Universe.  All classical
cosmological measurements depend on some combination of the expansion history
parameterized by $E(z)$ and the spatial curvature $R$ \cite{peebles}.  So, for
example, the linear growth of perturbations depends only on the expansion
history. The number-count--redshift relation, Alcock-Paczynski test, and
angular-diameter--redshift relation (determined, e.g., with radio sources)
depend on both the expansion and geometry, but none have been determined
precisely at sufficiently high redshift to tease out a cubic term.  We thus
conclude that the new supernova measurements are the first non-CMB measurements
to probe directly the effects of nontrivial spatial geometry.

\medskip

This work was supported at Caltech by NASA NAG5-11985 and DoE
DE-FG03-92-ER40701, and at Dartmouth by NSF grant PHY-0099543.

\end{document}